# Searching for a quantum critical point in YbCu$_{5-x}$Au$_x$

I. Čurlik[1], M. Reiffers[1,2], J. G. Sereni[3*], M. Giovannini[4], S. Gabani[1]

[1]*Institute of Experimental Physics, Watsonova 47, SK 043 53 Košice, Slovakia*
[2]*Faculty of Sciences, University of Prešov, 17. novembra 1, SK 080 78 Prešov, Slovakia*
[3] *Low Temperature Division, CAB-CNEA, 8400 San Carlos de Bariloche, Argentina*
[4]*CNR-SPIN and Department of Chemistry, University of Genova I-16146 Genova, Italy*

Abstract

Structural, magnetic, transport and thermal properties of YbCu$_{5-x}$Au$_x$ alloys with Au concentration between the limit of structural stability of AuBe$_5$ type at x = 0.4 up to x=0.7 are reported. The outstanding features of this system are: i) the constant and record high values of C$_m$/T ≈ 7J/molK$^2$ below a characteristic temperature T*, ranging between 150 mK and 350 mK. ii) A power law thermal dependence dependence C$_m$/T(T>T*) = A/T$^q$, with q = 1.3 ± 0.1, and iii) an arising incoherent electronic scattering observed in the resistivity at T< 1K for x ≤ 0.6 despite the fact that Yb magnetic atoms are placed in a lattice. Magnetic frustration, originated in the tetrahedral distribution of Yb atoms, appears as the responsible of the exotic behavior of this system.
    * Corresponding author: jsereni@cab.cnea.gov.ar

## I - Introduction

The strong correlation between electrons in lanthanide- and actinide based intermetallics gives rise to a variety of physical phenomena due to the hybridization of *f*-electrons and conduction electrons. The ground state (GS) formation of most of Ce, Yb and U compounds is driven by the competition between the inter-site magnetic interactions (RKKY), that favor magnetic order, and the local screening of *4f* magnetic moments induced by the Kondo effect [Stew84, Grewe90, Bau91, Ser98]. Since both effects are governed by the conduction- and *f*-electron exchange interaction (J$_{ex}$), the systems can be tuned by composition (i.e. the variation of their chemical potential) or by pressure from a magnetic to a non-magnetic behavior and vice-versa. Driving the magnetic phase boundary of the order transition to zero by means of these control parameters, the system may undergo quantum critical instabilities [HvL07]. Approaching that critical region, low energy collective excitations dominate the low temperature physical properties strongly rising the density of excitations as the temperature decreases. Logarithmic or power law temperature dependencies of their thermodynamic properties, like specific heat, magnetization or thermal expansion, reflect this scenario, recognized as non-Fermi-liquid behaviors [Stew01, Kuech03].

Among heavy fermion compounds, those which can be tuned close to a quantum critical point show the highest values of Cp/T for T→0, reaching up to 7J/mol K$^2$ for YbBiPt [Fisk91] and YbCo$_2$Zn$_{20}$ [Takeu11], 6.5 J/mol K$^2$ for PrInAg$_2$ [Yatskar96] and 5.6J/mol K$^2$ for CeNi$_9$Ge$_4$ [Killer04]. A common feature of these compounds is the contribution of their first excited crystal field levels to the GS properties because their respective Kondo temperatures are comparable to the crystal field splitting (i.e. T$_K$ ≈ Δ$_I$). For CePd$_3$B$_{0.5,}$ where the GS is a pure Kramer's doublet, similar high value of Cp/T(T→0) = 4.4 J/mol K$^2$ [Zeirin13] was found, concomitant with an extremely low characteristic temperature $T^* \approx 1K$, and reaching the full entropy $S_m = R\ln2$ of the GS at $T \approx 4K$ [Ser13a]. Finding other compounds which collect high density excitations at $T \to 0$ is highly desirable to provide further experimental evidences for this peculiar behavior,. particularly among Yb-based compounds with a pure Kramer's doublet GS.

For such a purpose, Yb compounds with very low temperature ($T \leq 1K$) magnetic transitions, tuned by alloying Yb ligand atoms merit, to be investigated, being $YbCu_{5-x}Au_x$ a proper candidate. $YbCu_4Au$ presents an extremely low Kondo temperature and it orders at $T < 1$ K [Galli02, Bau99]. This solid solution, crystallizing in the cubic $AuBe_5$-type structure, have recently received considerable attention due to the evolution of GS properties by substituting Au by Cu. The interest in this topic was initially triggered by the investigation on heavy fermions of the family $YbCu_4T$ (T = Ag, Au), which crystallize in the cubic $MgCu_4Sn$ type (an ordered derivative of the $AuBe_5$-type). In particular, it was found that $YbCu_4Au$ orders magnetically below 1 K [Ros87]. Later on, it was realized that $YbCu_4Au$, like analogous members of the type $YbCu_4T$ (T = Ag, In) [Yoshi97a, Michor02, He97], is actually a point of crystallographic order of a $YbCu_{5-x}Au_x$ solid solution. Starting from the magnetically ordered $YbCu_4Au$, the substitution of Au by Cu (i.e. decreasing x), drives the system to the disappearance of the magnetic order [Gio05]. On the stoichiometric limit, the non magnetic heavy fermion $YbCu_5$ (i.e. x = 0) could be prepared, as cubic $AuBe_5$, only at high pressure [Yoshi97b] or by melt spinning [Reif06] while careful structural analysis of $YbCu_{5-x}Au_x$ solid solutions [Carr03, Gio08] has shown that homogeneous compounds with cubic $AuBe_5$-type structure can form as single phase at ambient pressure only for $x \geq 0.4$, questioning the previous findings on the composition of a possible QCP below x = 0.4 [Yoshi01]. A transition temperature in the region close to the lower limit x = 0.4 has been determined just from the change in slope in the resistivity vs temperature [Yoshi01], whereas measurements of heat capacity have been done only for $x \geq 0.8$ [Galli02]. More recent measurements of zero-field and longitudinal field muon spin relaxations (μSR), nuclear quadrupole resonance (NQR), and magnetization measurements, seems to indicate the presence of ferromagnetic interactions at x = 0.6 [Carr09]. Thus, it still remains the open question about the eventual existence of a QCP and its position on concentration axis in $YbCu_{5-x}Au_x$ system. With the aim to attain more information on the evolution of the magnetic order at the low Au concentration limit and, consequently, trying to shed some light on the possible existence in the system of a QCP, the present paper reports our investigation of the $YbCu_{5-x}Au_x$ solid solution in the region close to the inferior limit, through measurements of magnetization above 2 K, and resistivity and heat capacity down to the mK region.

## II. Experimental details

The polycrystalline samples of $YbCu_{5-x}Au_x$ (x = 0.4, 0.5, 0.6, 0.7) have been prepared by weighting the stoichiometric amount of elements with the following nominal purity: Yb (99.9 pct mass), Cu (99.999 pct mass) and Au (99.99 pct mass). The elements were enclosed in small tantalum crucibles, sealed by arc welding under pure argon, in order to avoid the loss of Yb with a high vapor pressure. The samples were then melted in an induction furnace, under a stream of pure argon. To ensure homogeneity during the melting, the crucibles were continuously shaking. The samples were then annealed in a resistance furnace at 700 ºC for two weeks and finally quenched in cold water.

The alloys were characterized by optical and electronic microscopy and by quantitative electron probe microanalysis (EPMA). The crystalline structure was examined by X-ray diffraction (XRD). Heat capacity measurements in the temperature range 0.4 – 300 K and in an applicable magnetic field up to 9 T were performed by PPMS commercial device (Quantum Design) using the two-$\tau$ model of the relaxation method. For lower temperature range up to T = 0.05 K were used $^3$He-$^4$He dilution refrigerator. Electrical resistivity was measured also with PPMS device using 4-wire AC technique in 0.4 – 300 K temperature range. Magnetic susceptibility (at excitation field 50 Oe) and magnetization were measured by MPMS commercial device (Quantum Design) in the temperature range 2 – 300 K and in an applied magnetic field up to 5 T.

## III. Experimental Results

**A : Structural properties**

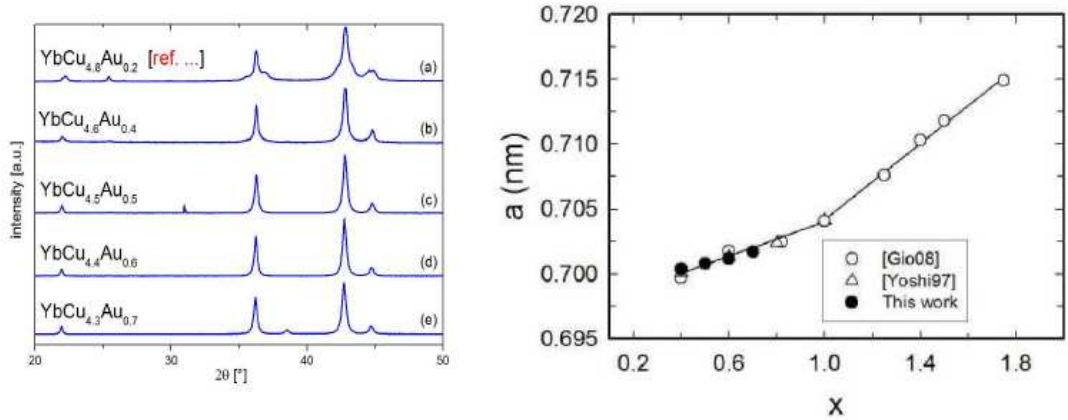

Fig. 1. X-ray diffraction (Cu K$_\alpha$) powder patterns for YbCu$_{5-x}$Au$_x$ alloys for (a) x = 0.2; (b) x = 0.4; (c) x = 0.5; (d) x = 0.6; (e) x = 0.7.
Fig. 2. Trend of the lattice parameter a for YbCu$_{5-x}$Au$_x$ compared with literature data

All the annealed samples remain cubic AuBe$_5$-type single phase as determined from EMPA and XRD. Fig. 1 reports XRD powder patterns for the four samples, together with data for x = 0.2 taken from our previous work [Gio08]. The patterns of the samples prepared in this work were indexed on the basis of the cubic AuBe$_5$-type. In particular, the satellite peaks of the monoclinic superstructure of the cubic AuBe$_5$-type, present for the alloy at x = 0.2, are absent in all the XRD patterns. Fig. 2 shows the variation of the lattice parameter *a* of the cubic AuBe$_5$ phase of YbCu$_{5-x}$Au$_x$ as a function of Au content compared with data taken from the literature. The values for the samples prepared in this work are in line with those reported in the literature [Yoshi01, Gio08]. As it was already evidenced in a previous paper [Gio08], the change of slope at the stoichiometric point of YbCu$_4$Au is due to the fact that YbCu4Au is at the crossover between two kinds of Cu/Au substituted sublattices, 16c for x > 1 and 4c for x < 1. Since these compounds crystalize in the AuBe$_5$ type structure, the Yb magnetic atoms are locii on a network of edge-sharing tetrahedra [Fritsch05]. The slope of $\delta a / \delta x$ for x < 1 is compatible with the atomic volume ratio between Au and Cu, therefore, no significant structural pressure is expected to be present in these alloys by decreasing Au content.

**B: Magnetic properties**

In Fig. 3 we display the thermal dependence of the magnetic susceptibility ($\chi$) measurements presented as the inverse $1/\chi$ within the range between 2 K and 300 K. These results show the typical Curie-Weiss law behavior for T > 50K, with an effective magnetic moment $\mu_{eff} \approx 4.3\ \mu_B$ as represented in the figure by a dashed line for comparison. This value is close to the expected for $Yb^{3+}$ ions with a $J = 7/2$ spin-orbit ground state, in good agreement with those found by Yoshimura et al. [Yoshi01]. Below 50K, the negative curvature indicates the effect of the thermal population reduction of the excited crystal electric field (CEF) levels. According to Lea Leask and Wolf (LLW) [LLW62], the CEF in cubic symmetry splits the eight fold $J = 7/2$ ground state into a two doublets ($\Gamma_6$ and $\Gamma_7$) and a quartet ($\Gamma_8$). Such a negative curvature in $1/\chi(T)$ reveals a reduction of $\mu_{eff}$ due a weaker intensity of the GS magnetic moment.

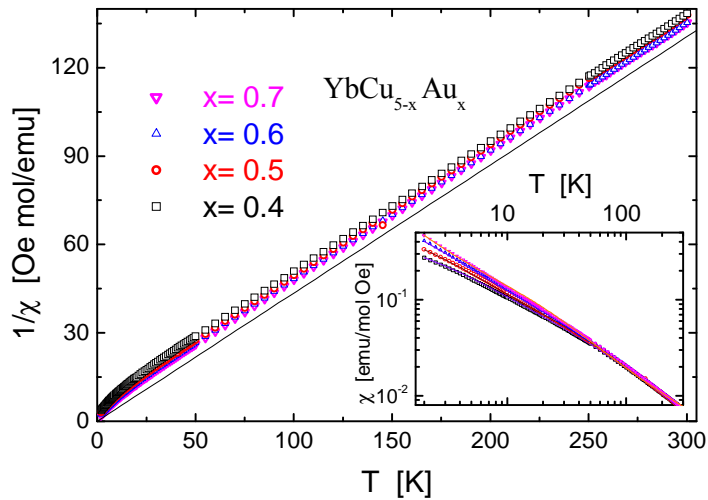

Fig. 3. Temperature dependence of the inverse magnetic susceptibility of $YbCu_{5-x}Au_x$ for x = 0.4, 0.5, 0.6, 0.7 in a field H = 50 Oe. Dashed line is a reference for a Curie law dependence with an effective magnetic moment $\mu_{eff}$ = 4.3 $\mu_B$. Inset: magnetic susceptibility in a double logarithmic representation to evidence the variation of the low temperature properties with Au concentration..

The contribution to the molecular field, manifested through the paramagnetic temperature ($\theta_P$), shows a concentration dependence between $\theta_P$ = -10K (for x = 0.7) and -18K (for x = 0.4), extrapolated from T > 50K. Its negative sign indicates an antiferromagnetic (AFM) character in the magnetic interaction. Also these values are in good agreement with the literature [Yoshi01]. The details of the low temperature $\chi(T < 50K)$ results are discussed in Section IV.

**C: Transport properties**

The temperature dependencies of the electrical resistivity ($\rho$) of YbCu$_{5-x}$Au$_x$ for $x$ = 0.4, 0.5, 0.6, 0.7 are shown in Fig. 4, compared with the $\rho(T)$ curves for x = 0.8 and 1 taken from the literature [Yoshi01], like those for 30mK < T < 1K included in the inset.

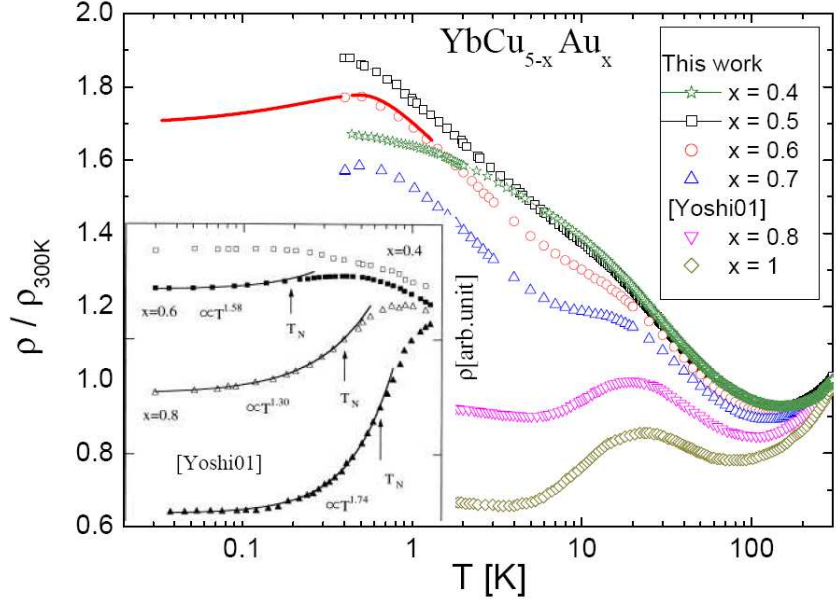

*Fig. 4. (Color online) Normalized thermal dependence of resistivity. Open symbols: results from YbCu$_{5-x}$Au$_x$ alloys, with x = 0.4, 0.5, 0.6 and 0.7 (this work) and x = 0.8, 1, after [Yoshi01]. Continuous curve (red): results down to 30 mK extracted from the inset. Inset: results down to 30 mK for x = 0.4, 0.6, 0.8, 1, after [Yoshi01].*

The high temperature region (above 30 K) of $\rho(T)$ of all the samples is characterized by a nearly logarithmic increase on decreasing the temperature, which is attributed to incoherent Kondo scattering related with the excited CEF levels. A local maximum around 20 K is ascribed to the position of the first crystal field level in the presence of Kondo interaction. However, marked differences exists between the ordered compound YbCu$_4$Au (x = 1) and the other alloys. In fact, YbCu$_4$Au exhibits typical Kondo lattice features with a clear onset of coherent scattering of the conduction electrons below about 20K. Conversely, the more disordered alloys (x = 0.7 and 0.6) are characterized by a progressive rise of the low temperature resistivity. Below 1K, an incipient flattening of $\rho(T)$ is observed for x ≤ 0.6, particularly in the x = 0.4 sample. Previous measurements down to 30mK [Yoshi01] show that the coherence effect strongly weakens at x = 0.6 and completely disappears at x = 0.4, see inset Fig. 4, where the incoherent scattering is shown. These features discard any possibility of magnetic order nor a Fermi-liquid GS formation as previously proposed [Carr09].

**D : Specific heat**

The specific heat of some these alloys was measured up to 60K, with the magnetic component ($C_m$) computed by subtracting the phonon contribution from the non magnetic isostructural compound YCu4Au [CurlikPhD]. In Fig.5 the high temperature thermal dependence of sample YbCu$_{4.3}$Au$_{0.7}$ is shown in a C$_m$ / T representation, within the range of temperature where the excited CEF levels develop their respective thermal dependent contributions. The inset displays the details of a fitting function that includes the corresponding Schottky anomalies for a doublet as first excited level and a quartet. Since a proper fit including usual hybridization effects (V$_{cf}$) between conduction and *4f* states acting on each excited level requires complex calculation protocols [Aligia13], a very simple criterion to mimic the levels broadening was applied in this case. For such a purpose each degenerated CEF level $\Gamma_i$, with respective degeneracies $\upsilon_i$ = 2 or 4, was split into single Dirac levels equally distributed in energy around the nominal value (i.e. the respective barycenter) of the original multiplet. This procedure requires the strength of the hybridization to be smaller than the CEF splitting ($\Delta_i$), i.e. V$_{cf}$ ~T$_K$ < $\Delta_i$, which is the case of this system since Mössbauer spectroscopy investigation reports an estimated hybridization scale of 0.3 K [Bonv92].

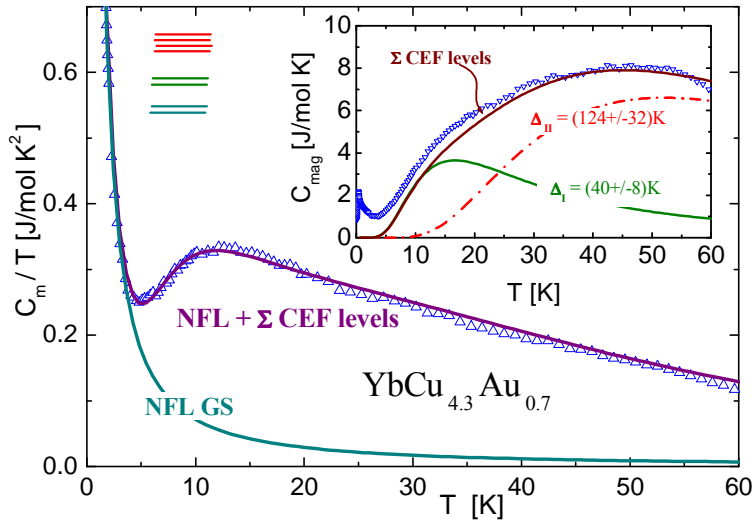

*Fig 5: High temperature magnetic specific heat divided temperature, showing the fit with the NFL behavior of the (doublet) GS and the contribution of the excited CEF levels: a doublet at $\Delta_I$ and a quartet at $\Delta_{II}$. Inset: detail of the contribution of each excited CEF (see the text).*

The fit was performed on sample x = 0.7 because it was measured up to T = 60K, see Fig. 5. Details of the fit accounting for the CEF levels contribution are included in the inset of that figure. The best description corresponds to two (Dirac) levels shifted ± 8 K from a center at $\Delta_I$ = 20 K, and two doublets shifted ± 32K from a center at $\Delta_{II}$ = 124K. The displacements in energy of those levels from their respective centers (i.e. from $\Delta_i$) can be taken as representative of the hybridization strength of each multiplet and consequently of their respective Kondo temperatures. With such a procedure, we obtain T$^I_K$ ~ 8K and T$^{II}_K$ ~ 32K respectively. One can check these values by applying Hanzawa's criterion: T$^h_K$ = (T$_K$ $\Delta_I$ .. $\Delta$n)$^{1/(n+1)}$ [Hanza85] and to evaluate the T$_K$ scale of the system as T$_K$ = (T$^h_K$)$^{(n+1)}$ / ($\Delta_I$ ...$\Delta$n). For n=1 (i.e. only including the first excited CEF level) one obtains T$_K$ = (T$^I_K$)$^2$ / $\Delta_I$ ≈ 1.6 K, and for n=3 (notice that the second excited state is a quartet) T$_K$ = (T$^{II}_K$)$^4$ / ($\Delta_I$ $\Delta_{II}^2$) = 1.7 K. Independently from the strict application of this model to our real system, it becomes clear that the scale of energy related to the Kondo effect is $T_K > 2K$. Similar values are obtained in section IV.2 from the analysis of thermal properties.

As it can be seen in Fig. 5, the sum of these Schottky anomalies plus the contribution of the GS (to be discussed below) provides a proper description of the experimental results including a simple representation of hybridization effects on each excited CEF level. Specific heat measurements cannot distinguish between two levels of the same degeneracy, in this case between the doublets of $\upsilon = 2$, $\Gamma_6$ and $\Gamma_7$, , whereas the $\Gamma_8$ quartet (with $\upsilon = 4$) becomes unambiguously identified as the highest excited level. From LLW both $\Gamma_6$ -$\Gamma_7$ -$\Gamma_8$ and $\Gamma_7$ -$\Gamma_6$ - $\Gamma_8$ spectral distributions are possible within the $-0.6 < x < 0.7$ range of the LLW-$x$ parameter, being the corresponding LLW-$w$ energy parameter determined as positive [LLW62].

Previous neutron scattering studies performed on YbCu$_4$Au report another CEF levels spectrum, where the quartet $\Gamma_8$ is the first excited CEF level [Sever90]. Such a level distribution is not consistent with the present $C_m(T)$ results because the maximum of the Schottky anomaly between the doublet GS and the quartet would exceed the measured Cm(T) values at $T = 0.42 \Delta_I$ . Nevertheless, the quasielastic line width ($\Gamma/2$) observed by neutron scattering is in good agreement with the $T_K$ extracted from the fits shown in Fig. 5 since $\Gamma/2 \approx 8$ K at $T = 20$ K and $\approx 22$ K at $T = 124$ K.

A relevant information for the understanding of the GS behavior extracted from this analysis is that the first excited CEF level practically does not contribute at low temperature because $T_K(V_{cf}) << \Delta_I$ . Thus, this system can be considered as the heaviest fermion reported at present with a pure doublet GS. Other mentioned Yb compounds with even higher $C_P/T$ ($T \rightarrow 0$) values, like those mentioned in the introduction, show more comparable values of $T_K$ and $\Delta_I$ .

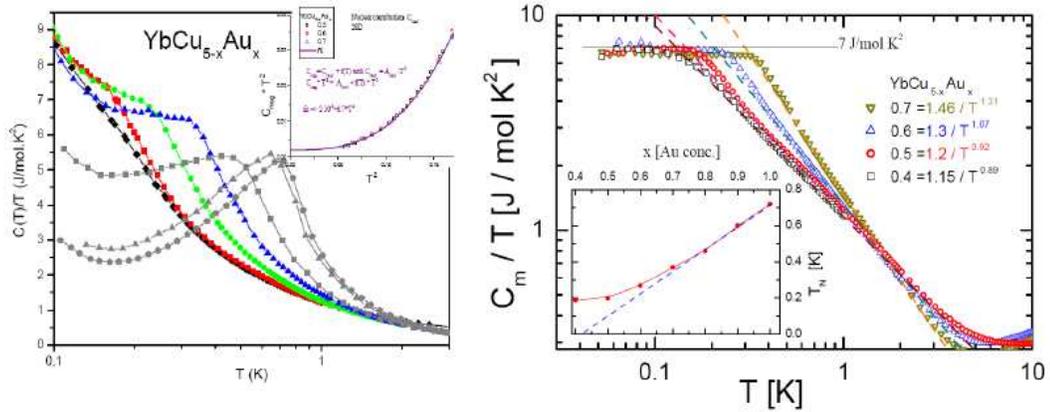

Fig. 6. Heat capacity plotted as Cp/T versus T on a logarithmic scale of YbCu$_{5-x}$Au$_x$ for $x = 0.4$, 0.5, 0.6, 0.7 (this work) and 0.8, 0.9, 1 after [Galli02]. Inset: Fit of the nuclear contribution in a $C_{meas} *T^2$ vs $T^2$ representation (see the text).

Fig. 7. (color online). Very low temperature magnetic contribution to specific heat in a double logarithmic representation. Dot lines represent the respective power law functions. Inset concentration dependence of $T_{max}.(x)$ split in two regions: as $T_N$ ($x \geq 0.8$) (blue squares) and $T^*$ ( $x \leq 0.7$) (red circles).

Low temperature specific heat measurements are displayed in Fig. 6 as $C_{meas}/T$ vs T in a semi-logarithmic scale for the studied YbCu$_{5-x}$Au$_x$ alloys with $x = 0.4$, 0.5, 0.6, 0.7, and complemented with data from the literature for x = 0.8, 0.9, 1 [Galli02]. YbCu$_4$Au exhibits a cusp in $C_{meas}/T$ around 0.7 K ascribed to the onset of long range magnetic order. At low temperatures (T< 0.4K), an upturn due to the nuclear contribution ($C_N$) increases the measured specific heat. Since that contribution decreases as $C_N = A_N / T^2$, in order to make a quantitative evaluation of that contribution we have fit the measured data using a $C_{meas} * T^2 = A_N + C_m(T) * T^2$ temperature dependence. From the fit performed on the studied samples up to T = 0.4K we obtained: $A_N = 2 * 10^{-3}$ and $C_m = \gamma_0 * T$, with $\gamma_0 = 6.7$ J/mol K$^2$, see the fit in the inset of Fig. 6. For practicality, we label the experimental $C_m/T$ (T→0) value as a $\gamma_0$ factor, nevertheless it has not to be identified with the standard Sommerfeld coefficient "$\gamma$" unless the low energy excitations are proved to form a (heavy/narrow) band. Such does not seem to be the case for this system because of the incoherent electronic scattering observed for the alloys with x ≤ 0.6.

As it can be observed in Fig. 6, the shape of the $C_m/T$ maximum ($C_m/T_{max}$) around T = $T_{max}$ changes progressively from a well defined cusp, for x ≥ 0.9 alloys, into a kink for x ≤ 0.7. This modification of the shape of the maximum coincides with a change in the value of $C_m/T_{max}$ from ~ 5 J/mol K$^2$ to ~6.7 J/mol K, suggesting some change in the nature of the GS. Within that range of concentration, $\gamma_0$ increases from 2.5 J/mol K$^2$ (at x = 0.9) up to a saturation value of 6.7 J/mol K$^2$ for x ≤ 0.7, whereas the $C_m/T \sim T$ dependence for T < $T_{max}$ is progressively replaced by a constant $C_m/T = \gamma_0$. One should notice the lack of $C_m(T)$ jumps defining a second order phase transition, instead a significant tail of $C_m/T$ above $T_{max}$ reveals an important fraction of degrees of freedom progressively condensed above that temperature. This feature is analyzed in section IV.4. in the context of the thermal entropy dependence.

## IV Discussion

**IV.1: Magnetic susceptibility**

In order to investigate the concentration effects on the low temperature magnetic properties of the alloys under study, in the inset of Fig.3 the experimental results of the χ(T) dependencies are presented in a double logarithmic representation. The experimental results are well described by a standard Curie-Weiss law including the CEF splittings and the respective Van Vleck contributions:

$\chi(T) = (g_J \mu_B)^2/Z^* [\Sigma_i |<\Gamma_n||\Gamma_n>|^2 e^{-\Delta/kT}/k_B(T+\theta_i) + \Sigma_{ij} |<\Gamma_n||\Gamma_m>|^2 (e^{-\Delta_i/kT} - e^{-\Delta_j/kT})/\Delta_{ij}]$

being Z the partition function [Lueken79]. As starting parameters, the CEF levels energies $\Delta_I = 40$ K and $\Delta_{II} = 124$ K, extracted from the high temperature $C_m$ measurements (see section III.D) are used. For simplicity, the respective level widths are not taken into account. This produces a minor deviation from the measurements at intermediate temperature, but not a low temperature because no broadening of the GS doublet is observed, in agreement with the mentioned Mössbauer results.

From the fits we obtain the effective paramagnetic moments: $\mu_{eff}$(GS) ≈ (2.8±0.3)$\mu_B$ and $\mu_{eff}(\Delta_I) = (3.7±0.3)\mu_B$. These values are very close to the expected theoretical values for a Yb$^{3+}$ ion with a $\Gamma_7$-$\Gamma_6$-$\Gamma_8$ spectral distribution [Birgen72] and converge to the J=7/2 value (4.54$\mu_B$) at high temperatures. The $\theta_{Pi}$ values obtained from T

> 50K results in subsection III.B for the CEF excited levels are in good agreement with those obtained with this fitting. However, a significant difference is found in the paramagnetic temperature computed at low temperature $\theta_P^{GS}(x)$ that runs between -0.7 K for x = 0.7 to -1.6 K for x = 0.4. These values indicate that any eventual Kondo screening is irrelevant for the GS properties within this range of Au concentration (i.e. $x \leq 0.7$). Notably, all the $\chi(T)$ results of these samples fit very well into the standard Curie-Weiss law [i.e. $\chi(T) \sim 1/(T+ \theta_P)$] with the only variation of $\theta_P^{GS}$, at variance from the $T^{(-2/3)}$ dependence reported in the literature [Carr09].

**IV.2: Specific Heat**

Fig. 7 collects the thermal dependencies of $C_m/T$ of the $0.4 \leq x \leq 0.7$ samples and, for comparison, those for $x \geq 0.8$ after Ref. [Galli02]. The former show a constant $C_m/T = \gamma_0$ plateau below a characteristic temperature T* ranging from $\approx$ 0.37K for x = 0.7 to $\approx$ 0.19K for x = 0.4 (see the insert in Fig.7). The observed value, $\gamma_0 \approx 6.7$J/mol K$^2$ is comparable to the highest values reported from some other Yb based compounds [Fisk91,Tori07] with $\gamma_0 \approx 7.3$J/mol K$^2$. Notably, the $\gamma_0(x)$ value increases with decreasing Au content from $\approx 1.5$J/mol K$^2$ for YbCu$_4$Au (i.e. x = 1) up to the reported value for x = 0.7 where it saturates. This variation is concomitant with the increase of the $C_m/T(x)$ maximum and the disappearance of the positive slope of $C_m/T$ below $T_{max}$. These features suggest a progressive transformation of the GS nature, from a phase that develops some type order parameter to some exotic phase with a constant density of excitations for $x \leq 0.7$. Remind that, for x = 0.6, no transition to a magnetically ordered phase was detected down to 20mK by μSR measurements [Carr09]. Instead, a dynamical muon relaxation driven by spin fluctuations and not by a static field distribution was found.

The low temperature plateau ends at a kink in the $C_m/T$ slope, at T = T*. Because of the clear change in the GS properties between x = 0.8 and 0.7, we label the temperature of the maximum of $C_m/T(T_{max})$ as $T_N$ on the region showing some kind of order parameter development, and as T* for $x \leq 0.7$ where $C_m/T(T\to 0)$ is constant. In the latter region, the mentioned plateau of $C_m/T$ is followed above T = T* by a power law thermal dependence of the type: $C_m/T = B/T^Q$, with B ranging between 1.43 and 1.22 J/mol K$^{1+Q}$ and Q between 1.3 and 0.95 for x = 0.7 and x = 0.4 respectively. This thermal dependence of $C_m/T$ also changes for $x \geq 0.8$ as a further confirmation of the intrinsic transformation of the GS nature. In the inset of Fig. 7, the phase boundary represented by $T_N(x)$ and T*(x) is presented.

In comparison with the reported interpretation of resistivity measurements [Yoshi01] we have seen that this system shows a continuous transformation between x = 0.9 and 0.7, with the possibility of the mixture of two components suggested by the progressive increase of $\gamma_0(x)$ between those Au concentrations. The thermal dependency of the electrical resistivity [Yoshi01] behaves accordingly, because no kinks in ρ(T) are observed to allow to define $T_N(x)$, instead it is defined by the onset of coherence in the scattering of the conduction electrons with the magnetic lattice (see the inset in Fig. 4). The effect of coherence practically disappears at x = 0.6 and for x = 0.4, where ρ(T) increases continuously by cooling indicating the presence of incoherent scattering like in a single impurity behavior. Nevertheless, the present ρ(T) measurements on the x = 0.4 sample show a more pronounced flattening at low temperature respect to the x = 0.5 one. This effect may be attributed to a reduction of disorder as Au concentration decreases, and to better sample quality respect to the one of the literature. This phenomenology was also observed in other very heavy fermion systems ($\gamma_0 > 4$J/mol K$^2$) like CePd$_3$B$_{0.6}$ [Nieva88] and CeNi$_9$Ge$_4$ [Killer04], suggesting that a very high density of excitations does not imply the formation of an electronic band with coherent character.

Coming back to the phase boundary, one sees that between $1 \geq x \geq 0.8$, the concentration dependence of $T_N(x)$, defined by the cusp in $C_m/T$ and the onset of coherence in $\rho(T)$ [Yoshi01] extrapolates to zero around x = 0.42. However, for $x \leq 0.7$ the temperature of the kink in $C_m/T$ at $T^*(x)$ tends to saturate a finite value ($T^* \approx 0.1$K) as *x* decreases, as shown in the inset of Fig. 7. This change in the concentration dependence between $T_N$ and $T^*$ coincide with the changes mentioned before in the $C_m/T(T)$ properties..The presence of a plateau is an unusual thermal dependence, only observed in the few heavy fermions showing the upper limit $\gamma_0$ values [Ser13b]. The question arises whether this peculiar feature is related to the extremely high values of $C_m/T(T)$ reached by the divergent power law thermal dependence above $T^*$. It is evident that in case to keep diverging at low temperature it would have exceeded the entropy of a doublet GS (i.e. Rln2). Notice that the Rln2 constraint for the two fold degenerated GS implies that to such a high values of $\gamma_0$ corresponds a very low energy scale, i.e. a very low characteristic temperature. This condition excludes the $T_K$ values of $\approx 10$K proposed in the literature [Yoshi01].

**IV.3 Magnetic frustration**

These peculiar spectroscopic and thermodynamic results, including the observed dynamical features [Carr09], converge into the phenomenology of frustrated magnetic system with signs of spin liquid behavior. In fact, this structural configuration realizes the optimal conditions for a three dimensional frustration because, as already mentioned in subsection III.A, in the AuBe$_5$-type structure the magnetic Yb atoms are located on a network of edge-sharing tetrahedra. A further evidence for such a scenario is currently given by the empirical frustration parameter f = |$\theta_P$| / $T_N$ [Ramirez94] that, for $0.7 \geq x \geq 0.4$ alloys, rises from $\approx 1.9$ up to $\approx 8.4$.

An illustrative comparison can be done with the isotypic compound GdCu$_4$In [Fritsch05] and the Cd doped alloys [Fritsch06]. In this case, In substitution by Cd enhances next-nearest-neighbor (NNN) magnetic interactions allowing them to compete with next-neighbor (NN) ones. As a consequence a relaxation of the frustration conditions occurs [Fritsch05]. In the case of YbCu$_{5-x}$Au$_x$, Au and Cu atoms are isoelectronic, but their different atomic size introduces some atomic disorder. Although the increase of incoherent scattering observed from stoichimetric YbCu$_4$Au (see subsection III.C) may be partially explained by a growing atomic disorder, the mentioned NQR studies [Carr09] revealed a dynamic character the spin susceptibility. Such a behavior cannot be attributed to static atomic disorder only but, more likely, to some frustration effect.

The unusual appearance of a plateau in $C_m/T$ below $T^*$ for $x \leq 0.7$, which replaces the maximum observed at $x \geq 0.8$, also indicates that some microscopic mechanism inhibits even the formation of any sort of short range order. This feature also points to the presence of magnetic frustration effects that arise by proximity to the ideal edge-sharing tetrahedral network of Yb atoms in the ideal $YbCu_5$ limit [Bau95, Reif04, Mito06], where 4*c* and 16*e* crystalline sites would be simultaneously occupied by Cu atoms.

A further test of this scenario can be performed by studying the effect of applied magnetic field, since above a threshold one may expect to quench spin fluctuations. As it can be seen in Fig. 8, the magnetic field *H* on sample x = 0.4 reduces progressively the density of low energy excitations, in coincidence with the nuclear spin relaxation rates results [Carr09].

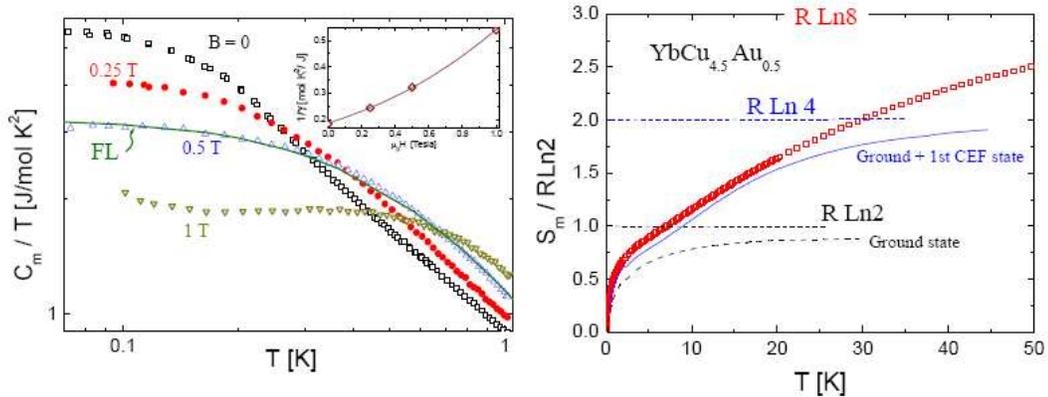

*Fig. 8: (Color online) Effect of the magnetic field on the specific heat in sample x = 0.4 in a double logarithmic representation. Full curve represents a Fermi liquid thermal dependence. Inset:. inverse of $\gamma_0$ as a function of applied field.*

*Fig. 9: (color online) Thermal variation of the entropy of $YbCu_{4.5}Au_{0.5}$ up to 50K. Dash-lines are references for the entropy of the excited CEF levels. Lower dash-curve represents the entropy of the GS., while the upper continuous-curve indicates the sum of the ground and the first CEF doublet, both computed from the thermal dependencies shown in Fig.5 .*

However, the observed $1/\gamma_0$ dependence with *H* is linear for *H* <0.5T (see inset Fig. 8), which is not on line with the conclusion that a Fermi liquid (FL) state is induced by magnetic field because, in such, a case a $1/\gamma_0$ vs $H^2$ dependence is predicted [Edelst88]. It is only under an applied field of 0.5T that a FL thermal dependence is reached for the specific heat, as depicted by the full curve in Fig. 8. Therefore, this alloys behave as a non-FL in agreement with the observed power law thermal dependence above T = T*. Nevertheless, the spin fluctuation that control the GS properties are dumped by a field of ≈ 0.5T. Notice that the energy $\mu_B H$ associated to an applied magnetic of 0.5 T compares with the thermal energy $k_B T^*$ .

**IV.4: Entropy**

In Fig. 9 the thermal variation of the magnetic entropy ($S_m$) of $YbCu_{4.5}Au_{0.5}$ up to 50K is presented. There, $S_m(T)$ shows a clear tendency to reach the total value of R*ln*8 expected for a *J* = 7/2 Hund's rule ground state above the temperature range of our measurements. The R*ln*2 value corresponding to a doublet GS is not completely reached because of the magnetic frustration at low temperature. Nevertheless, the value of $S_m(T)$ at T ≈ 6 K guaranties that only a doublet is involved in the GS state properties. his is in agreement with the upturn of $C_m/T$ (T) around 5K (see Fig. 5) where the first excited CEF doublet (at $\Delta_I$ ≈ 40K) starts to contribute to the specific heat. In Fig.9, the lower (black) dash-curve represents the entropy of the GS, while the upper (blue) continuous-curve indicates the sum of the ground and the first CEF levels, both computed from the thermal dependencies shown in Fig.5

In order to characterize the low temperature behavior from this entropy information, one can evaluate a scale of energy $T_0$ applying the criteria used for Kondo impurities proposed by Desgranges-Schotte for a doublet GS within the single ion Kondo model [Desgr82]: i) $T_0 = (\pi R/3) / \gamma_0$ that, with a $\gamma_0 = 6.7 J/molK^2$ value $T_0 \approx 1.2$ K and ii) the temperature at which $S_{mag}= 2/3\, Rln2$ (see Fig.9) corresponds to $T_0 = 1.6$ K. Both values are in very good agreement with those obtained previously from the analysis on the high temperature specific heat of $T_0 \sim 1.5$ K (section III.D). These $T_K$ values are notably small in comparison with those observed in other Yb and Ce based intermetallic compounds. The present scenario can be understood by the vicinity of this alloys to a quantum critical point where there are the conditions for a possible Kondo break down as proposed some years ago [Q-Si06]. As mentioned before, from the basic thermodynamic constraint related to the number of degrees of freedom of a doublet GS is fix (c.f. total $S_m = Rln2$), one confirms that to very high $\gamma_0$ values corresponds the obtained very small characteristic scales of energy.

## V. Conclusions

In this work, a systematic investigation on a Yb-lattice with record high value of $C_m /T(T \to 0) \approx 6.7 J/molK^2$ is presented. Although such a value compares with a few previous cases reported in the literature, in this case corresponds to a pure Kramer's doublet ground state as proved by high temperature specific heat analysis. Such a large density of excitations at low temperature appears as an empirical upper limit because, once reached after decreasing the Au content from x = 1 to x = 0.7, it remains constant. That a limit is imposed by the amount of the available degrees of freedom fixed by $S_m = Rln2$. Above a characteristic temperature T* the $C_m /T(T)$ plateau transforms into a power law temperature dependence without evidence of any $C_m(T)$ jump.

The apparent contradiction between the arising incoherent electronic scattering observed in the resistivity at T < 1K for x ≤ 0.6 despite the fact that Yb magnetic atoms are placed in a lattice is explained by a magnetic frustration, originated in the tetrahedral distribution of Yb atoms placed in the '4a' sites. Dynamical muon relaxation driven by spin fluctuations confirm these findings.

## References

[Bau91] E. Bauer, Adv. in Phys. 40 (1991) 417.
[Bau95] E. Bauer, Le Tuan, R. Hauser, E. Gratz, T. Holubar, G. Hilsher, K Yoshimura, J.Magn. Magn. Mat., 140-144 (1995) 1247.
[Bau99] E. Bauer et al. Phys.Rev. B 60 (1999) 1238.

[Bau05] E. Bauer, G. Hilscher, H. Michor, C. Paul, Y. Aoki, H. Sato, DT Adroja, JG Park, P. Bonville, C. Godart, J. Sereni, M. Giovannini, A. Saccone, *J. Phys. Cond. Mat.* 17 S999-S1009 (2005).
[Birgen72] R.J. Birgeneau; J. Phys. Chem. Solids 33 (1972)59.
[Bonv92] P. Bonville, B. Canaud, J. Hammann, J.A. Hodges, P. Imbert, G. Jéhanno, A. Serering, Z. Fisk; J. Phys, I France 2 (1992) 459.
[Carr03] P. Carretta, M. Giovannini, M. Horvatic, N. Papinutto, A. Rigamonti, Phys. Rev. B Rapid Comm. 68  220404-1, 2204404-4 (2003).
[Carr09] P. Carretta, R. Pasero, M. Giovannini, C. Baines, Phys.Rev.B 79 (2009) R020401
[CurlikPhD]. I. Curlik, PhD Thesis, Univerzita Pavla Jozefa Šafárika v Košiciach, 2012.
[Sever90] A. Severing, P. Murani, J.D. Thompson, Z. Fisk, C.-K. Loong, Phys. Rev. B 41 (1990) 1739.
[Desgr82] H.-U. Desgranges and K. D. Schotte, Phys. Lett. 91A (1982) 240
[Aligia13]  M.A. Romero, A.A. Aligia, G.L. Nieva and J.G. Sereni; J. Phys. Condens Matter 26 (2014) 025602.
[Edelst88]. A.S. Edelstein; Phys. rev. B 37 (1988) 3808.[Fritsch05] V. Fritsch, J.D. Thompson, J.L. Serrao; Phys. Rev. B 71 (2005) 132401.
[Fisk91] Z. Fisk, P.C. Canfield, W.P. Beyermann, J.D. Thompson, M.F. Hundley, H.R. Ott, E. Felder; Phys. Rev. Lett. 67 (1991) 3310.
[Fritsch05] V. Fritsch, J.D. Thompson, J.L. Serrao, Phys. Rev. B 71 (2005) 132401.
[Fritsch06] V. Fritsch, J.D. Thompson, J.L. Serrao, H.-A. Krug von Nidda, R.M. Eremina, A. Loidl; Phys. Rev. B 73 (2006) 094413.
[Galli02]  M. Galli,, E. Bauer, St. Berger, Ch. Dusek, M. Della Mea, H. Michor, D. Kaczorowski, E.W. Scheidt, F. Marabelli, Physica B 312-313 (2002) 489-491.
[Gio2000] M. Giovannini, H. Michor, E. Bauer, G. Hilscher, P. Rogl, T. Bonelli, F. Fauth, P. Fischer, T. Hermannsdorfer, L. Keller, W. Sikora, A. Saccone, R. Ferro, *Phys.Rev.B* 61, 4044 (2000).
[Gio03]  M. Giovannini, A. Saccone, P. Rogl, R.Ferro, *Intermetallics* 11 (2003) 197.
[Gio05] M. Giovannini, A. Saccone, St. Muller, H. Michor, E. Bauer, *J. Phys: Condens. Matter* 17, S877 (2005).
[Gio08] M. Giovannini, R. Pasero, S. De Negri, A. Saccone, *Intermetallics* 16, 399 (2008).
[Grewe90] N. Grewe and F. Steglich, in *Handbook of Physics and Chemistry of Rare Earths*, Eds. L. Eyring and K. Gschneidner Jr., Elsevier Pub. 1990, Vol 14, Ch. 98.
[Hanza85] K. Hanzawa, K. Yamada, K. Yosida; J. Magn. Magn. Mat. 47&48 (1985) 357.
[He97] J.He, N.Tsujii, K.Yoshimura, K.Kosuge, T.Goto, J.Phys.Soc.Jap.66 (1997) 2481-6.
[HvL07] H. von Löhneysen, A. Rosch, M. Vojta, and P. Wölfle, in *Fermi-liquid instabilities at magnetic quantum phase transitions;* Rev. Mod. Phys. 79, 1015 (2007) or cond-mat/0606317.
[Killer04] U. Killer, E.-W. Scheidt, G. Eickerling, H. Michor, J. Sereni, Th. Pruschke, S. Kehrein; *Phys. Rev. Lett.* 92, 27003 (2004).
[Kuech03] R. Küchler, N. Oeschler, P. Gegenwart, T. Cichorek, K. Neumaier, O. Tegus, C. Geibel, J. A. Mydosh, F. Steglich, L. Zhu, Q. Si; Phys. Rev. Lett. 91, (2003) 066405.
[Lueken79] see for ejample: H. Lueken, W. Brüggemann, W. Bronger, J. Fleischhauer, J. Less. Common Metals 65 (1979) 79-88.
[LLW62]  K.R. Lea, M.J. Leask, W.P. Wolf; J. Phys. Chem. Solids 23 (1962) 1381.

[Michor02] H. Michor, K Kreiner, N. Tsuji, K. Yoshimura, K. Kosuge, G. Hilscher Physica B 319 (2002) 277-81.
[Mito06] T.Mito, M. Nakamura, M. Shimoide, M. Otani, T. Koyama, S. Wada, H. Kotegawa, T.C. Kobarashi, B. Idzikowski, M. Reiffers, J.L. Serrao, Physica B 379-380 (2006) 732.
[Mur11] T. Muramatsu, T. Kanemasa, T. Kagayama, K. Shimizu, Y. Aoki, H. Sato, M. Giovannini, P. Bonville, V. Zlatic, I. Aviani, R. Khasanov, C. Rusu, A. Amato, K. Mydeen, M. Nicklas, H. Michor, E. Bauer, Phys. Rev. B 83 (2011) 180404(R).
[Mydosh93] J.A. Mydosh, in *Spin Glasses*, Taylor & Francis, 1993.
[Nieva88] G.L. Nieva, PhD Thesis, Universidad Nacional de Cuyo, 1988.
[Q-Si06] Q. Si, Physica B 378-380 (2006) 23
[Ramirez94] A.P. Ramirez, Ann. Res. Matter. Sci. 24 (1994) 453.
[Reif04] M. Reiffers, B. Idzikowski, S. Ilkoviĉ, A. Zorkovská, J. Šebek, K.H. Müller, J. Magn. Magn. Mat. 272-276 (2004) 209.
[Reif06] M. Reiffers, B. Idzikowski, J. Sebek, E. Santava, S. Ilkovic, G. Pristas Physica B 378-380, 738 (2006).
[Ros87] C. Rossel, K.N. Yang, M.B. Maple, Z: Fisk, E. Zirgnabl and J.D. Thompson *Phys. Rev.* B 35 (1987) 1914–8.
[Ser98] J.G. Sereni, J. Phys. Soc. Jpn. 67 (1998)) 1767.
[Ser13a] J.G. Sereni, G. Schmerber, J.P. Kappler; IEEE Trans. Magnetics. 49 (2013) 4647.
[Ser13b] J.G. Sereni; Phil. Mag. 93 (2013) 409-433;
[Stew84] G.R. Stewart, in *Heavy-fermion systems*., Rev. Mod. Phys. 56,:755 (1984)..
[Stew01] G.R. Steward, in *Non-Fermi Liquids;* Rev. Mod. Phys. *73, 797–855 (2001).*
[Takeu11] T. Takeuchi et al.; Journal of Physics: Conference Series 273 (2011) 012059
[Tori07] M. S. Torikachvili S. Jia, E. D. Mun, S. T. Hannahs, R. C. Black, W. K. Neils, D. Martien, S. L. Bud'ko, P.C. Canfield; PNAS, 104 (2007) 9960
[Yatskar96] A. Yatskar, W.P. Beyermann, R. Movshovich, P.C. Canfield; Phys. Rev. Lett. 77 (1996) 3637.
[Yoshi97a] K. Yoshimura, N. Tsujii, J. He, M. Kato, K. Kosuge, H. Michor, K. Kreiner, G. Hilscher, T. Goto J. Alloys Compounds 262-263 (1997) 118-23.
[Yoshi97b] K. Yoshimura, N. Tsujii, J. He, M. Kato, K. Kosuge, H. Michor, et al. J. Alloys Comp 262-263 (1997) 11.
[Yoshi01] K. Yoshimura, T. Kawabata, N. Sato, N. Tsujii, T. Terashima, C. Terakura, G. Kido, and K. Kosuge, J. Alloys Compd**.** 317-318 (2001) 465.
[Zeirin13] I. Zeiringer, J.G. Sereni, M. G.-Berisso, K. Yubuta, P. Rogl, A. Grytsiv, E. Bauer; Mater. Res. Express 1 (2014) 016101.